# LLR-based Successive-Cancellation List Decoder for Polar Codes with Multi-bit Decision

Bo Yuan and Keshab K. Parhi, *Fellow, IEEE*

*Abstract*—Due to their capacity-achieving property, polar codes have become one of the most attractive channel codes. To date, the successive cancellation list (SCL) decoding algorithm is the primary approach that can guarantee outstanding error-correcting performance of polar codes. However, the hardware designs of the original SCL decoder have large silicon area and long decoding latency. Although some recent efforts can reduce either the area or latency of SCL decoders, these two metrics still cannot be optimized at the same time. This paper, for the first time, proposes a general log-likelihood-ratio (LLR)-based SCL decoding algorithm with multi-bit decision. This new algorithm, referred as *LLR-$2^K$b-SCL*, can determine $2^K$ bits simultaneously for arbitrary $K$ with the use of LLR messages. In addition, a reduced-data-width scheme is presented to reduce the critical path of the sorting block. Then, based on the proposed algorithm, a VLSI architecture of the new SCL decoder is developed. Synthesis results show that for an example (1024, 512) polar code with list size 4, the proposed LLR-$2^K$b-SCL decoders achieve significant reduction in both area and latency as compared to prior works. As a result, the hardware efficiency of the proposed designs with *K*=2 and 3 are 2.33 times and 3.32 times of that of the state-of-the-art works, respectively.

## I. Introduction

POLAR codes [1] have emerged as one of the most attractive forward error correction (FEC) codes in recent years. Due to their unique capacity-achieving property, polar codes provide outstanding error-correcting capability that would be very useful for digital transmission.

However, to date polar codes suffer from inferior finite length error-correcting performance. In particular, in the region of short or medium code length, polar codes are not comparable to the LDPC codes in terms of coding gain. To solve this problem, successive-cancellation list (SCL) decoding algorithm was proposed in [2] to improve the coding gain of the polar codes. In [2], it was shown that with the use of SCL algorithm, polar codes can outperform the WiMAX LDPC codes even for a shorter code-length.

Although SCL algorithm can help polar codes achieve beyond-LDPC performance, this approach suffers from high complexity and long latency. Recently, some efforts were proposed to address these problems. An LLR-based SCL algorithm was proposed in [3-4] to reduce the amount of combinational logic and memory. In [5-6][11], low-latency SCL algorithms were presented to reduce the required number of decoding cycles. However, these prior works only focused on either reducing area or latency, but not on optimizing these two metrics at the same time.

This paper, for the first time, proposes a general reduced-latency LLR-based SCL decoding algorithm. This new algorithm, namely *LLR-$2^K$b-SCL*, can determine $2^K$ bits in one cycle for arbitrary *K* with the use of LLR messages. As a result, it can achieve both low complexity and short latency. In addition, a reduced-data-width scheme is presented to reduce the critical path of the sorting block. Based on the proposed algorithm, a VLSI architecture of the new SCL decoder is developed. Synthesis results show that for an example (1024, 512) polar code with list size 4, the proposed LLR-$2^K$b-SCL decoder achieves great reduction in both area and latency as compared to the prior works, respectively. As a result, the hardware efficiency of the proposed designs with *K*=2 and 3 are 2.33 times and 3.32 times of that of the state-of-the-art works, respectively. Notice that this paper is a generalized version of our prior work [10].

The rest of the paper is organized as below. Section II gives a brief review of polar codes. The proposed LLR-$2^K$b-SCL algorithm and the reduced-bit-width scheme for sorting block are presented in Section III. Section IV develops the hardware architecture of the new algorithm. Hardware performance is analyzed and discussed in Section V. Section VI draws the conclusions.

## II. Review of Polar Codes

### A. Encoding of Polar Codes

The encoding procedure of (*n*, *p*) polar codes consists of two steps. First, the *p*-bit source message is extended to an *n*-bit message $\boldsymbol{u}=(u_1, u_2,…u_n)$ with padding *n-p* "0" bits. Here those padded "0" bits are referred as *frozen bits* and their positions over the $\boldsymbol{u}$, namely *frozen positions*, are known to both the transmitter and receiver. Then, $\boldsymbol{u}$ is multiplied with an *n*-by-*n* generator matrix $\boldsymbol{G}$ to obtain the transmitted codeword $\boldsymbol{x}=(x_1, x_2,…,x_n)=\boldsymbol{uG}$. Fig. 1 shows the example of an *n*=4 polar codes encoder. For details on encoding of polar codes, the reader is referred to [1].

Manuscript received June 19$^{th}$, 2015.
Bo Yuan was with Department of Electrical and Computer Engineering, University of Minnesota; he is now with Department of Electrical Engineering, City University of New York, City College. Email: byuan@ccny.cuny.edu
Keshab K. Parhi is with Department of Electrical and Computer Engineering, University of Minnesota. Email: parhi@umn.edu.

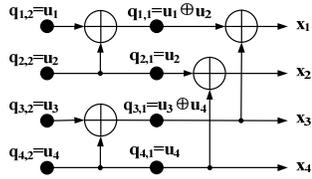

Fig. 1. *n*=4 polar encoder, cited from [5].

*B. Successive-Cancellation (SC) Decoding Algorithm*

At the receiver end, the original transmitted codeword *x* is corrupted by noise to the received codeword $y=(y_1, y_2,…,y_n)$. An LLR-based successive-cancellation (SC) decoder can be used to recover *u* from *y*. Fig. 2 shows an example SC decoding procedure for *n*=4 polar codes. It can be seen that the SC decoder consists of two types of computation units, namely **f** and **g**, respectively. Here each **f** or **g** unit is labelled with a number, which indicates the time index when the corresponding unit is activated for computation. It can be seen that at clock cycle 2, 3, 5 and 6, the **f** or **g** unit in the last stage (stage 2 in this example) sends its LLR output to the hard-decision **h** unit to determine the decoded bit. Notice that the example SC decoder is based on LLR form. In this case, the function of **f** and **g** units can be approximated as (1) and (2):

$$f(a, b)=\text{sign}(a)\text{sign}(b)\min(|a|, |b|) \quad (1)$$

$$g(a,b) = a(-1)^{\hat{u}_{sum}} + b \quad (2)$$

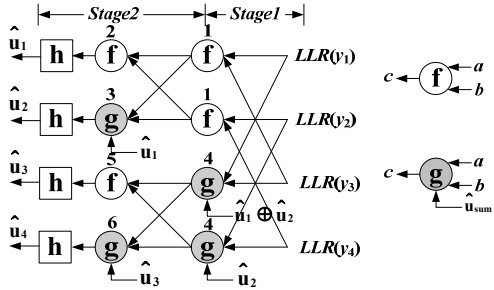

Fig. 2. Example LLR-based SC decoding scheme for *n*=4, cited from [7].

*C. Successive-Cancellation List (SCL) Algorithm*

The decoding process of polar codes can also be interpreted from the view of code tree. Fig. 3 shows an example code tree for *n*=4 polar codes. Here each level represents a decoded bit. The value that is associated with each node represents the probability (*metric*) of the corresponding decoding path. For example, 0.23 at level 4 represents the probability for the length-4 path (1000) is Pr($\hat{u}_1$=1, $\hat{u}_2$=0, $\hat{u}_3$=0, $\hat{u}_4$=0)=0.23. Based on this representation, the objective of a successful decoding procedure is to find the length-*n* path that is the corrected codeword. To achieve this goal, the SC decoder first visits the children nodes that are associated with the current survival path at each level. Then it selects the new path that has the larger metric as the updated survival path. Because this searching strategy is only locally optimal, the performance of the SC decoder is limited.

Different from the SC decoder that only selects a single path, the *L*-size SCL algorithm utilizes *L* different searching paths. Therefore, it is more likely for the SCL algorithm to find the desired path than the SC algorithm. For example, SCL decoder with *L*=2 is able to trace the valid path (1000) in Fig. 3 while the SC decoder fails to find it.

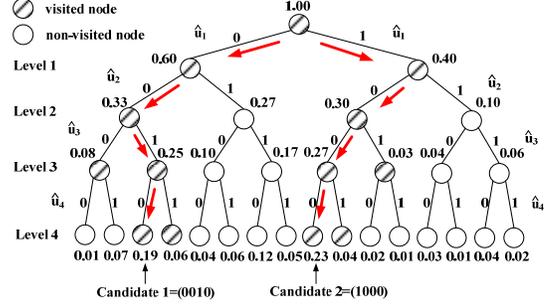

Fig. 3. Searching of SCL decoder with *n*=4, *p*=4 and *L*=2, cited from [5].

### III. THE PROPOSED LLR-BASED MULTI-BIT DECODING

From Fig. 3 it can be seen that the SCL algorithm needs to visit the nodes at each level of the code tree. This traversing searching style leads to a very long decoding latency. To address this problem, a reduced-latency algorithm that can determine $2^K$ bits simultaneously was proposed in [5]. However, the approach in [5] is based on the likelihood messages, which requires much larger computation and memory complexity than the commonly used LLR-based decoder in [3-4]. On the other hand, those LLR-based SCL decoders in [3-4] are only able to determine one bit in one cycle; hence they have much longer latency than the decoder in [5]. As a result, to date those prior SCL decoders in [3-5] are not able to reduce the latency and area at the same time.

*A. LLR-based Multi-bit SCL Decoding*

This subsection presents an LLR-based SCL decoding algorithm with $2^K$ bits decision, namely *LLR-$2^K$b-SCL*. Notice that [5] also proposed a reduced-latency decoder that can determine multiple bits in one cycle. However, the SCL decoder in [5] is based on the likelihood form while the design in this paper is based on the LLR form. This difference in the fundamental representation of the proposed algorithm leads to much less computation complexity than the approach in [5].

Next, we show how to determine each successive $2^K$ bits as $\hat{u}_{2^K(i-1)+1}$, …, $\hat{u}_{2^K i}$ at the same time in the LLR form. From the view of code tree (see Fig. 3), this means the SCL decoder is able to directly calculate the metrics of length-($2^K i$) paths from the metrics of length-($2^K(i-1)$) paths. In general, such direct computation is performed by an LLR-based *metric computation unit* (MCU), which replaces the original last *K* stages of each LLR-based SC decoder (see Fig. 2). In the following paragraph, we show how to derive the function of the LLR-based MCU.

Assume that the previously decoded $2^K(i-1)$ bits $\hat{u}_1$, …, $\hat{u}_{2^K(i-1)}$ are $z_1$, …, $z_{2^K(i-1)}$, respectively. This event is denoted as $\hat{u}_1^{2^K(i-1)} = z_1^{2^K(i-1)}$. Therefore, in the logarithmic domain the length-($2^K i$) path metric can be represented as:

$$M(\alpha_1…\alpha_{2^K}, z_1^{2^K(i-1)}) \triangleq \ln(\Pr(\hat{u}_{2^K(i-1)+1}^{2^K i} = \alpha_1^{2^K}, \hat{u}_1^{2^K(i-1)} = z_1^{2^K(i-1)})), \quad (3)$$

where $\hat{u}_{2^K(i-1)+1}^{2^K i}$ is defined as ($\hat{u}_{2^K(i-1)+1}$, …, $\hat{u}_{2^K i}$) that is the set of the current $2^K$ decoded bits. In addition, $\alpha_1^{2^K}$ is defined as ($\alpha_1$, …, $\alpha_1^{2^K}$) whose elements are the binary values.

(3) contains the probabilistic information of the current $2^K$ decoded bits, which is unknown during the decoding procedure. To address this problem, we need to further represent the logarithmic path metrics with the LLR messages that are input to the MCU. Such reformulation is based on the fact that the polar decoding procedure is inherently "guided" by its encoding procedure [8]. Simultaneous right-to-left decoding procedure of the successive $2^K$ bits (see Fig. 4(a)) involves the estimation of the left-to-right encoding procedure (see Fig. 4(b)). Hence if $\hat{u}_{2^K(i-1)+1}^{2^K i}$ is estimated to be $\alpha_1^{2^K}$, then $\widehat{out}_1^{2^K} \triangleq (\widehat{out}_1, \ldots, \widehat{out}_{2^K})$ should be the estimation of $\alpha_1^{2^K} U$, where $U$ is the $2^K$-by-$2^K$ generator matrix. As a result, (3) can be further re-written as:

$$M(\alpha_1\ldots\alpha_{2^K}, z_1^{2^K(i-1)}) = \ln(\Pr(\widehat{out}_1^{2^K} = \alpha_1^{2^K} U, \hat{u}_1^{2^K(i-1)} = z_1^{2^{K(i-1)}})) \quad (4)$$

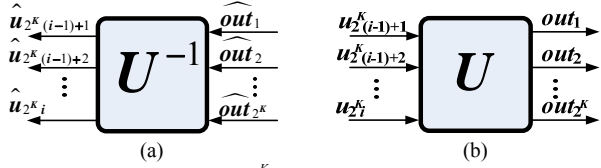

Fig. 4. (a) right-to-left decoding of $2^K$ bits. (b) left-to-right encoding bits.

Notice that the determination of $\widehat{out}_1, \ldots, \widehat{out}_{2^K}$ are independent. In addition, if we denote the $j$-th column vector of $U$ as $U(j)$, then we have $\widehat{out}_j = \alpha_1^{2^K} U(j)$. As a result, (4) can be further derived as below:

$$\begin{aligned}&M(\alpha_1\ldots\alpha_{2^K}, z_1^{2^K(i-1)}) \\&= \ln(\Pr(\widehat{out}_1^{2^K} = \alpha_1^{2^K} U \mid \hat{u}_1^{2^K(i-1)} = z_1^{2^{K(i-1)}})\Pr(\hat{u}_1^{2^K(i-1)} = z_1^{2^{K(i-1)}})) \\&= \sum_{j=1}^{2^K} \ln(\Pr(\widehat{out}_j = \alpha_1^{2^K} U(j) \mid \hat{u}_1^{2^K(i-1)} = z_1^{2^{K(i-1)}})) + M(z_1^{2^K(i-1)}),\end{aligned} \quad (5)$$

where $M(z_1^{2^K(i-1)}) = \ln(\Pr(\hat{u}_1^{2^K(i-1)} = z_1^{2^{K(i-1)}}))$ is the logarithmic length-($2^K(i-1)$) path metric.

Recall that each SC component decoder is based on LLR form. In that case, the $j$-th input to the MCU block is:

$$s_j = \ln(\frac{\Pr(\widehat{out}_j = 0 \mid \hat{u}_1^{2^K(i-1)} = z_1^{2^K(i-1)})}{\Pr(\widehat{out}_j = 1 \mid \hat{u}_1^{2^K(i-1)} = z_1^{2^K(i-1)})})$$

As a result, we can obtain the elements of the first item in (5):

$$\Pr(\widehat{out}_j = 0 \mid \hat{u}_1^{2^K(i-1)} = z_1^{2^K(i-1)}) = \frac{e^{s_j}}{e^{s_j}+1}$$
$$\Pr(\widehat{out}_j = 1 \mid \hat{u}_1^{2^K(i-1)} = z_1^{2^K(i-1)}) = \frac{1}{e^{s_j}+1} \quad (6)$$

Substituting (6) into (5), we have:

$$M(\alpha_1\ldots\alpha_{2^K}, z_1^{2^K(i-1)}) = \sum_{j=1}^{2^K}(s_j(1-\alpha_1^{2^K}U(j)) - \ln(e^{s_j}+1)) + M(z_1^{2^K(i-1)}) \quad (7)$$

(7) describes the LLR-based update principle for path metrics. Once the MCU block receives the $2^K$ input LLR messages $s_j$ and the previous metric of length-($2^K(i-1)$) path, it can immediately calculate the new metric of length-($2^K i$) path with the use of (7), which corresponds to the simultaneous decision for $2^K$ bits.

Notice that (7) contains exponential and logarithmic functions, which require long critical paths in hardware design. Therefore, (7) needs to be simplified for feasible VLSI implementation.

Consider $\ln(1+e^x) \approx x$ for large $x$; otherwise 0 for small $x$. (7) can be further approximated as below:

$$M(\alpha_1\ldots\alpha_{2^K}, z_1^{2^K(i-1)}) \approx \sum_{j=1}^{2^K}(s_j(1-\alpha_1^{2^K}U(j)) - \delta(s_j)) + M(z_1^{2^K(i-1)}), \quad (8)$$

where $\delta(s_j)=s_j$ if $s_j \geq 0$; otherwise 0.

(8) shows how to directly calculate the metric of length-($2^K i$) paths from the metric of length-($2^K(i-1)$) paths. With the use of this update principle, we can develop the LLR-based SCL decoding algorithm with $2^K$ bits decision as Scheme-A. In general, an $L$-size LLR-$2^K$b-SCL decoder consists of $L$ copies of LLR-based SC decoder. To decode every $2^K$ successive bits, each SC component decoder first performs the regular SC decoding procedure till the last-($m$-$K$) stage (see Fig. 2), where $m=\log_2 n$. At this time, the ($m$-$K$) stage outputs $2^K$ LLR messages $s_j$ ($j=1, 2,\ldots 2^K$) to the MCU block. Then, the MCU block in each SC component decoder calculates the new path metrics with the use of (8). After that, all of the updated path metrics from the $L$ SC component decoders are compared and $L$ largest are selected as the survival paths metrics. The above entire procedure is repeated for every $2^K$ bits until all the $n$ bits are determined. Note that similar to [5], a simple zero-forcing unit (ZFU) is needed after the computation of (8), which helps to drop the unqualified paths that violate the frozen conditions.

---
**Scheme A: L-size LLR-$2^K$b-SCL Algorithm for (n, k) polar codes**
---
1: **Input:** *Log - Likihood ratios of each bit in the received codeword*
2: **Initialization:** *Path metric $M_0 = 0$ for each survival path*
3: **For** $i = 1$ **to** $n/2^K$
4:    **For each length-($2^K(i-1)$) survival path** $\hat{u}_1^{2^K(i-1)} = z_1^{2i-2}$
5:    **SC component decoding:**
6:        Activate stage -1 to stage -($m$-$K$) of LLR-based SC decoder
7:        stage -($m$-$K$) output $2^K$ LLR-based message $s_j$ ($j=1,2,\ldots,2^K$)
8:    **Path Expansion:**
9:        Expand survival path $z_1^{2^K(i-1)}$ to $2^{2^K}$ candidate paths ($\hat{u}_1^{2^K i} = (\alpha_1^{2^K}, z_1^{2i-2})$)
10:        $1$ length -($2^K(i-1)$) path $\Rightarrow 2^{2^K}$ length -($2^K i$) paths
11:    **Metric Computation:**
12:        Calculate $2^{2^K}$ actual path metrics $M(\alpha_1\ldots\alpha_{2^K}, z_1^{2^K(i-1)})$ by (8)
13:    **Forcing Zero:**
14:        $M(\alpha_1\ldots\alpha_{2^K}, z_1^{2^K i})$ for path $\hat{u}_1^{2^K i} = (\alpha_1^{2^K}, z_1^{2i-2})$
15:        $\hat{u}_{2^K(i-1)+j}$ is frozen $\Rightarrow$ all $M(\alpha_1\alpha_2\ldots\alpha_{j-1}0\alpha_{j+1}\ldots\alpha_{2^K}, z_1^{2^K(i-1)}) = -\inf$
16:    **End for**
17:    **Compare and Prune:**
18:        Compare $M(\alpha_1\ldots\alpha_{2^K}, z_1^{2^K(i-1)})$ for all the $2^{2^K}$ length -($2^K i$) candidate paths
19:        Select $L$ paths with the $L$ largest metrics as the new survival paths
20: **End for**
21: **Output:** *Choose the length - n survival path with the largest metric*
---

### B. Reduced-Data-Width Scheme for Sorting Block

Typically, $Q=6$ bit quantization scheme is sufficient for the fixed-point implementations of LLR-based SCL decoder. However, as indicated in [3], the representation of path metrics needs more bits since the path metrics have larger data range than the propagated LLR messages. In [3], it showed that $M=8$ bit for path metrics can avoid significant performance degradation in terms of frame error rate (FER). Since the overall critical path of the SCL decoder is in the sorting block that sorts those path metrics [3][5], the escalating data-width of path metrics inevitably causes significant increase in criticial path delay. To address this challenge, we propose a reduced-bit-width scheme for sorting block. The key idea is to only utilize $S=M-1$ bits to represent the path metrics for sorting, while the representations of path metrics for updating and storing are still based on $M$ bits. This approach is derived from the following observation: In SCL decoder the sorting block does not require as high precision as MCUs, since the function of the soring block is just to rank the path metrics without changing their values, while the MCUs have to adopt larger data-width to guarantee accurate calculation. Therefore, a reduced-data-width for sorting block does not cause significant performance degradation but enables reduction in critical path delay. Fig. 5 shows the fixed-point simulation results for the proposed LLR-$2^K$b-SCL algorithm with reduced-data-with for sorting block. Here the simulation environment is AWGN channel with BPSK modulation and the code parameters are $n=1024$, $r=0.5$. From the figure it can be seen that, compared to the original scheme using $S=8$ bits for sorting the path metric, the proposed scheme with $S=7$ bits only has negligible performance loss for different values of $K$ and $L$.

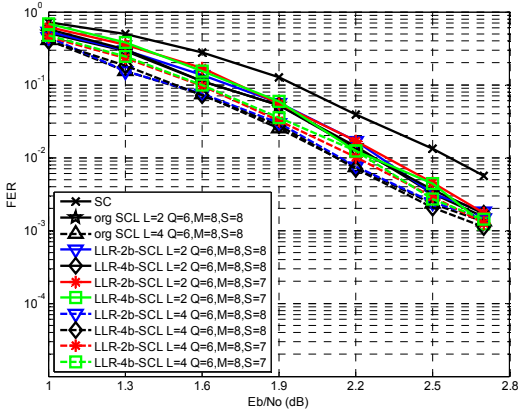

Fig. 5. Simulation results of (1024, 512) polar codes over AWGN channel.

## IV. Hardware Architecture

### A. Overall Architecture

Fig. 6 shows the overall hardware architecture of the proposed $L$-size LLR-$2^K$b-SCL algorithm decoder. The data path of the entire decoder contains $L$ LLR-based SC decoders plus a metric sorting block. For each SC component decoder, it is reformulated from the LLR-based decoder in [8], which retains the ($m$-$K$) stages but the last $K$ stages are replaced by the LLR-based MCU and ZFU. In addition, the required memory resource of the entire decoder consists of register arrays, bulk memory and buffer for survival paths, path metrics, propagating LLR messages and channel outputs, respectively.

In this section, the hardware design of the sorting block is very straightforward and similar to the approaches in [3] [5]. The only difference is that the data-width of comparators is reduced from $M$ bits to $M-1$ bits and the least significant bits (LSBs) of all the input path metrics are dropped to be consistent with the data-width of comparators. Therefore, in this section we focus on other parts of the data path and memory resource. Notice that since ZFU can be easily implemented with multiplexers; the analysis of this block is omitted as well.

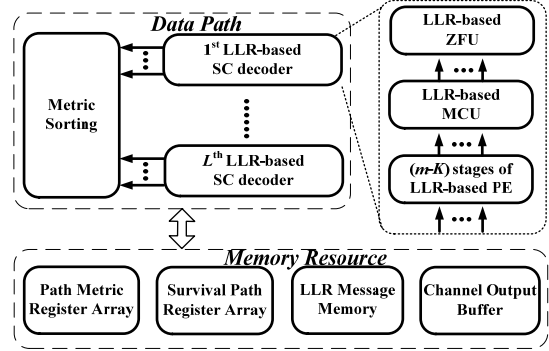

Fig. 6. Overall architecture of the LLR-$2^K$b-SCL decoder.

### B. LLR-based Metric Computation Unit (MCU)

In Section III (8) describes the function of MCU. Since this function depends on $K$, the hardware design of MCU varies with different choices of $K$. Fig. 7 illustrates the inner architecture of MCU for $K=3$. Here $\delta(\cdot)$ block can be simply implemented with a multiplexer. In addition, StoC and CtoS blocks represent the components that perform the conversion between sign-magnitude and 2's complement forms.

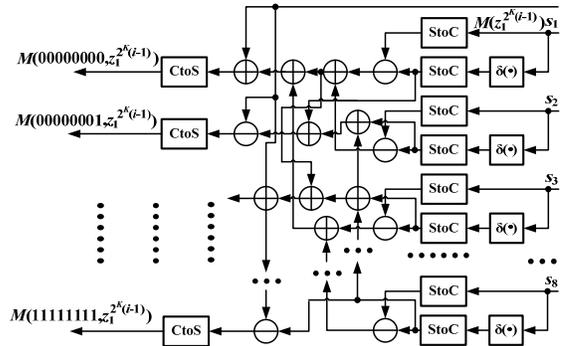

Fig. 7. Architecture of MCU for $K=3$.

### C. LLR-based Processing Element (PE)

As shown in line 5 − line 7 in Scheme-A, the input LLR messages of MCU are calculated from the first ($m$-$K$) stages of LLR-based SC decoder in [8]. In general, these stages consist of **f** and **g** nodes in Fig. 2, which can be implemented in hardware as the following processing element (PE) in Fig. 8. Notice here the addition and subtraction in (2) is designed as a unified adder and subtractor to save area.

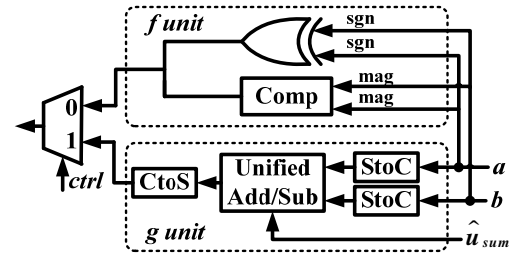

Fig. 8. Architecture of PE.

### D. Memory Resource

For the proposed decoder, different types of memory resource are needed for different types of data. Because the $L$ survival paths and their metrics need to be updated simultaneously each time, they are stored in registers. In addition, for the propagating LLR messages that are processed in the PEs, $L$ bulk memory banks are used to store the corresponding messages in the $L$ different SC component decoders. This memory partition approach can also avoid the potential memory access conflict. Besides, a specific buffer is used to store the initial LLR inputs to the decoder.

## V. Performance and Discussion

### A. Hardware Performance Comparison

Table I shows the hardware performance of different SCL decoders with list size $L=4$ for the same (1024, 512) polar codes. For the designs with different technology libraries, the results on area and frequency are scaled to the same 90nm nodes. From this table it can be seen that; the proposed design with $K=3$ can achieve much higher throughput than the same design with $K=2$ with a slight increase in area and critical path delay (lower clock frequency). As a result, the hardware efficiency of the design with $K=3$ can achieve 572Mbps/mm$^2$, which is the highest among the listed works.

TABLE I. Hardware Performance of (1024, 512) Decoders with $L=4$.

| Design | This work[††] | | [3] | [5] | [9][†] |
|---|---|---|---|---|---|
| Message form | LLR | | LLR | LL | LL |
| # of PE/path | 64 | | 64 | 1023 | 64 |
| Multi-bit decision | Yes | | No | Yes | No |
| Tech. (nm) | 65 | | 90 | 65 | 90 |
| Quantization. | 6-bit | | 6-bit | Dynamic | Dynamic |
| K | 2 | 3 | N/A | 2 | N/A |
| Area (mm$^2$) | 0.94[*] | 1.18[*] | 1.78 | 4.10[*] | 2.46 |
| Freq. (MHz) | 390[*] | 360[*] | 794 | 288[*] | 492 |
| Latency (Clock cycle) | 1056 | 546 | 2648 | 1022 | 2590 |
| Throughput (Mbps) | 378 | 675 | 307 | 288 | 194 |
| Efficiency (Mbps/mm$^2$) | 402 | 572 | 172 | 70 | 79 |

[*] The results have been scaled to 90nm.
[†] Re-synthesis results for 64PE of each path are cited from [3].
[††] The compensation operation for permutation matrix $B_N$ has been embedded in the design of the proposed decoders.

Compared with the LL-based designs [5] and [9], the proposed design with $K=3$ has 71% and 52% less area, respectively. In addition, its low decoding latency leads to 31% and 94.8% increase in throughput, respectively.

Compared with the LLR-SCL decoder in [3], the proposed design with $K=3$ has 79.3% shorter latency, which translates to 133% increase in decoding throughput. It should be noted that the proposed works with $K=2$ and 3 adopt data path balancing technique in [5] to reduce the critical path delay. If advanced technique, such as optimized sorting block in [3] is utilized, the clock frequency of the proposed designs will be further improved, thereby leading to even higher throughput and hardware efficiency.

### B. Discussion on relevant works

In [3], a LLR-based SCL decoder was proposed. The derivation for the LLR representation in [3] was similar to the work in [4] with slight difference on the sign of path metrics. However, the bit-decision in [3][4] is serial, thereby causing low throughput. Different from these works, this paper enables the simultaneous decision of each $2^K$ bits with the LLR representation; hence it can achieve much higher throughput and lower latency. For instance, as seen in Table I, the proposed design with $K=3$ achieves 133% increase in data rate than [3].

In [9], a channel message compression scheme was proposed to reduce the memory requirement. However, because [9] is only an LL-based decoder, its silicon area is at least two times of the proposed design. Interestingly, because the area-optimizing techniques in [9] and this paper are performed at different levels, they can be jointly used to develop a more area-efficient decoder.

To address the long latency problem, multi-bit decision, or so-called parallel output was proposed in [5-6][11]. These literatures describe the reduced-latency technique in different manners, but all utilize the special recursive property of polar codes. However, different from those prior LL-based works, the proposed approach successfully enables multi-bit decision with LLR-based messages, thereby leading to great reduction in both computation complexity and memory requirement, which are very important for application of polar codes.

## VI. Conclusion

In this paper we present LLR-$2^K$b-SCL algorithm for polar codes decoding. The proposed algorithm can reduce complexity and decoding latency at the same time without performance loss. Then, based on the proposed algorithm, we develop the corresponding VLSI architecture. Hardware analysis shows that the proposed SCL decoders have significant reduction in area and decoding latency.

## References

[1] E. Arıkan, "Channel polarization: A method for constructing capacity-achieving codes for symmetric binary-input memoryless channels," *IEEE Trans. Inf. Theory*, vol. 55, no. 7, pp. 3051-3073, 2009.

[2] I. Tal and A. Vardy, "List decoding of polar codes," arXiv:1206.0050, May 2012.

[3] A. Balatsoukas-Stimming, M. Bastani Parizi and A. Burg, "LLR-based successive cancellation list decoding of polar codes,"arXiv:1401.3753v3.

[4] B. Yuan and K.K. Parhi, "Successive cancellation list polar decoder using Log-likelihood ratios," in *Proc. of Asilomar Conf. on Signal, Systems and Computers*, pp. 548-552, 2014.

[5] B. Yuan and K.K. Parhi, "Low-latency successive-cancellation list decoders for polar codes with multi-bit decision," *IEEE Trans. on VLSI Systems*, vol. 23, no. 10, pp. 2268 – 2280, Oct. 2015.

[6] B. Li, H. Shen, D. Tse and W. Tong, "Low-Latency Polar Codes via Hybrid Decoding," in *Proc. of 8$^{th}$ Intl. Symp. on Turbo Codes and Iterative Info. Processing (ISTC)*, pp. 223-227, Aug. 2014.

[7] B. Yuan and K.K. Parhi, "Successive cancellation decoding of polar codes using stochastic computing," in *Proc. of IEEE Intl. Symp. on Circuits and Systems (ISCAS)*, May 2015.

[8] B. Yuan and K.K. Parhi, "Low-Latency successive-cancellation polar decoder architectures using 2-bit decoding," *IEEE Trans. Circuits and Systems-I: Regular Papers*, vol. 61, no. 4, pp. 1241-1254, Apr. 2014.

[9] J. Lin and Z. Yan, "An efficient list decoder architecture for polar codes," accepted by *IEEE Trans. on VLSI Systems*, 2015.

[10] B. Yuan and K.K. Parhi, "Reduced-latency LLR-based SC list decoder for polar codes," in *Proc. of 2015 ACM Great Lakes Symposium on VLSI*, pp. 107-110, May 2015.

[11] H. Vangala, E. Viterbo and Y. Hong, "A New Multiple Folded Successive Cancellation Decoder for Polar Codes," in *Proc. of IEEE Information Theory Workshop (ITW)*, pp. 381-385, Nov. 2014.